\documentclass[12pt]{article}
\newif\iffigs\figstrue

\usepackage{epsfig,latexsym}
\usepackage{amsmath}
\usepackage{color}
\usepackage{graphicx}
\usepackage{verbatim}
\usepackage{mathrsfs}
\usepackage{amssymb}

\DeclareMathAlphabet{\mathpzc}{OT1}{pzc}{m}{it}

 \csname
@addtoreset\endcsname{equation}{section}

\def\gz0{\gamma^{0}}

 \def\det{{\rm det\,}}

\def\m{\mu}

\def\bec{\begin{center}}
\def\ec{\end{center}}

\def\12{\frac{1}{2}}

\def\DH{\rm I\kern-1.5pt\rm H\kern-1.5pt\rm I}

\def\DR{\rm I\kern-1.45pt\rm R}
\def\DC{\kern2pt {\hbox{\sqi I}}\kern-4.2pt\rm C}

\newcommand{\I}{{\rm i}}

\newcommand{\beq}{\begin{equation}}
\newcommand{\eeq}{\end{equation}}
\newcommand{\bea}{\begin{eqnarray}}
\newcommand{\eea}{\end{eqnarray}}
\newcommand{\bi}{\begin{itemize}}
\newcommand{\ei}{\end{itemize}}

\usepackage{slashed}

\usepackage{float}

\usepackage{caption}

\usepackage{epsf}
\usepackage{amsmath}

\usepackage{multirow}

\usepackage{epsfig}
\usepackage{cite}
\usepackage{color,colordvi}

\newcounter{hran}

\makeatletter
\renewcommand\section{\@startsection {section}{1}{\z@}%
                               {-3.5ex \@plus -1ex \@minus -.2ex}%
                               {2.3ex \@plus.2ex}%
                               {\normalfont\large\bfseries}}
\makeatother

\vspace{0.5cm}

\begin{document}
\thispagestyle{empty}

\begin{flushleft}
\hskip 12cm CERN-TH-2017-131 \\
\hskip 12cm CPHT-RR043.062017 \\
\end{flushleft}

\vspace{15pt}

\begin{center}


{\Large\sc A superfield constraint for ${\cal N}=2 \to {\cal N}=0$ breaking }\\


\vspace{35pt}
{\sc E.~Dudas${}^{\; a}$, S.~Ferrara${}^{\; b,c,d}$ and A.~Sagnotti${}^{\; e}$}\\[15pt]

{${}^a$\sl\small Centre de Physique Th\'eorique, \'Ecole Polytechnique and CNRS \\ 91128 Palaiseau \ FRANCE
\\ }e-mail: {\small \it
emilian.dudas@cpht.polytechnique.fr}\vspace{6pt}

{${}^b$\sl\small Department of Theoretical Physics\\
CH - 1211 Geneva 23, SWITZERLAND \\ }
e-mail: {\small \it sergio.ferrara@cern.ch}
\vspace{6pt}

{${}^c$\sl\small INFN - Laboratori Nazionali di Frascati \\
Via Enrico Fermi 40, I-00044 Frascati, ITALY}\vspace{6pt}

{${}^d$\sl\small Department of Physics and Astronomy, Mani L. Bhaumik Institute for Theoretical Physics \\ U.C.L.A., Los Angeles CA 90095-1547, USA}\vspace{6pt}

{${}^e$\sl\small
Scuola Normale Superiore and INFN\\
Piazza dei Cavalieri \ 7\\I-56126 Pisa \ ITALY \\}
e-mail: {\small \it sagnotti@sns.it}\vspace{6pt}

\vspace{8pt}

\vspace{25pt} {\sc\large Abstract}
\end{center}

\noindent
We identify a cubic holomorphic constraint that subtends the total breaking of ${\cal N}=2$ supersymmetry in a vector multiplet and exhibit its microscopic origin. The new constraint leaves behind, at low energies, a vector and the two goldstini, in a non--linear Lagrangian that generalizes the ${\cal N}=2$ Volkov--Akulov model.
\vfill
\noindent

\baselineskip=20pt

\newpage

\setcounter{page}{2}
\setcounter{equation}{0}
\section{Introduction and Summary} \label{sec:intro}

Nonlinear realizations of supersymmetry via constrained superfields \cite{va,rocek} have been recently the focus of a wide activity, since the convenient off--shell formulations proposed in \cite{nonlinear,ks} have paved a way to a wide range of generalizations~\cite{generalizations} and applications.  The latter include supergravity models of the inflationary phase~\cite{vasugra} (see~\cite{fks} for a recent review), the coupling of the Volkov--Akulov model to supergravity, in its dual (higher--derivative~\cite{DFKS}  and two--derivative~\cite{BKVP}) forms, toy models of orientifold vacua~\cite{orientifolds} with ``brane supersymmetry breaking''~\cite{bsb} and detailed realizations of the KKLT construction~\cite{KKLT}, whose potential uplift emerges from similar aspects of D-brane dynamics.

It is well known that one can build an ${\cal N}=2$ vector multiplet combining an ${\cal N}=1$ vector multiplet $V=(A_\m,\lambda,D)$ and an ${\cal N}=1$ chiral multiplet $\Phi=(Z,\psi,F)$ \cite{WB}, while also attaining an off--shell formulation. One can thus address a number of interesting questions in a relatively handy way, in particular working in terms of ${\cal N}=1$ superfields. One of the supersymmetries is then manifest in superspace, while the other mixes the two superfields according to
\bea
& \delta_2 \, \Phi & = \ i\, \sqrt{2}\,\epsilon_2^\alpha\, W_\alpha \ , \nonumber \\
& \delta_2\, W_\alpha & = \ - \ i\,\sqrt{2} \left(m \ -\ \frac{1}{4} \ \overline{D}^{\,2} \,\overline{\Phi}\right)\epsilon_{2\alpha} \ + \  \sqrt{2}\, \sigma^\mu_{\alpha\dot{\alpha}}\, \partial_\mu \Phi \ {\overline{\epsilon}_2}^{\dot{\alpha}} \ . \label{second_SUSY}
\eea
Here $m$ is the ``magnetic'' parameter of~\cite{APT}, whose role in the partial breaking of supersymmetry \cite{HP} was clarified in~\cite{APT,FGP}, and following common practice we are resorting to the two--component formalism in the notation of \cite{WB}.

The vacuum shifts of the two Fermi fields follow from eqs.~\eqref{second_SUSY} and from their ${\cal N}=1$ superfield counterparts, and are encoded in the complex two-by-two matrix
\beq
{\cal M} \ =  \ \left( \begin{array}{cc}
    \sqrt{2}\,F     &  i\, D  \\
  i \,D      &  \sqrt{2}\,\left({\overline F}+m\right)
\end{array} \right)\ , \label{matrixMitro}
\eeq
so that the transformations at stake read
\beq
\begin{pmatrix}
    \delta \psi     \\
    \delta \lambda
\end{pmatrix}  =
\begin{pmatrix}
    \sqrt{2}\, F      & i D  \\
    i\, D       &  \sqrt{2} ({\overline F} \ +  \ m)
\end{pmatrix}
\begin{pmatrix}
    \epsilon_1   \\
    \epsilon_2
\end{pmatrix}  \ + \  \cdots \ .
\eeq
Notice also that
\bea
{\rm tr} \left( {\cal M}^{\dagger}\, {\cal M} \right) &=& 2\bigl[\left|F\right|^2 \, +\, \left|F+m\right|^2 \,+\,D^2\bigr] \ , \label{MdagMintro1} \\
\det \left( {\cal M}^{\dagger} \, {\cal M} \right) &=& \bigl|2\,F \left({\overline F} \ + \ m\right) \ + \ D^2 \bigr|^2  \ . \label{MdagMintro2}
\eea
Let us anticipate that, up to a normalization, eq.~\eqref{MdagMintro1} defines the ${\cal N}=2$ vacuum energy determined by the supersymmetry Ward identities~\cite{APT,adft}, while the determinant in eq.~\eqref{MdagMintro2} defines a scale
\beq
\Lambda \ = \ \bigl|D^2 \ + \ 2\,F({\overline F}\,+\, m)\bigr|^\frac{1}{4} \ , \label{scale}
\eeq
which will play a central role in the ensuing discussion and vanishes in the vacuum if at least one supersymmetry is unbroken. Clearly, if both supersymmetries are unbroken, the trace in eq.~\eqref{MdagMintro1} also vanishes in the vacuum. Moreover, if at least one supersymmetry is unbroken, the ${\cal N}=2$ vacuum energy differs in general from the ${\cal N}=1$ vacuum energy computed from superspace $F$ and $D$-terms~\cite{APT,adft}. The eigenvalues of ${\cal M}^{\dagger}\, {\cal M}$ are of the form
\beq
V_{\pm} \ = \ V \ \pm \ \left| \mu \right| \ ,
\eeq
where the trace determines the ${\cal N}=2$ vacuum energy, and if $V=|\mu|$ one of the supersymmetries is unbroken.

The non--linear limit of the partial ${\cal N}=2 \to {\cal N}=1$ breaking was studied long ago in \cite{BG}, where it was connected to the ${\cal N}=1$ supersymmetric Born--Infeld theory \cite{BI,DP,CF}. Important additions to the picture can be found in a number of subsequent works, and in particular in \cite{RT,BMZ,KT}. For the ensuing discussion, the highlight of these developments is the quadratic constraint
\beq
{\cal C}_2: \ W^\alpha\, W_\alpha \ - \ 2\, \Phi \left( \frac{1}{4} \ \overline{D}^{\,2}\, \overline{\Phi} \ - \ m \right) \ = \ 0 \ , \label{2_1_constraint}
\eeq
which was identified in \cite{BG} and encodes the non--linear realization of the second supersymmetry. Eq.~\eqref{2_1_constraint} expresses all components of the chiral multiplet $\Phi$ in terms of the vector multiplet, and consistently with their elimination has two notable consequences,
\beq
\Phi^2 \ = \ 0  \qquad {\rm and} \qquad \Phi\, W_\alpha \ = \ 0 \ , \label{21_consequences}
\eeq
to which we shall return shortly.

In view of the ensuing discussion, it is important to stress that the two conditions in eq.~(\ref{21_consequences}), the first of which also entered the early construction in~\cite{nonlinear}, have different types of solutions. A first type of solution leaves no residual supersymmetry and eliminates the gaugino in terms of the Volkov-Akulov goldstino
\cite{ks}, which is left over together with the gauge field. In contrast, a second type of solution determines $\Phi$ in terms of $W_\alpha$ and ${\overline W}_{\dot \alpha}$. Only if one demands that the second nonlinear supersymmetry of eq.~(\ref{second_SUSY}) be present, however, does the solution become unique, with the goldstino that completes the ${\cal N} = 1$ vector multiplet.  On the other hand, without the second supersymmetry  eqs.~(\ref{21_consequences}) admit an infinite class of ${\cal N}=1$ supersymmetric solutions. The simplest is perhaps $\Phi = W^2$, which adds to the supersymmetric Maxwell theory a quartic interaction arising from the D-density of the squared supercurrent.

As shown in \cite{STORA}, the constraint of eq.~\eqref{2_1_constraint} affords a natural generalization to the case of several Abelian vector multiplets~\footnote{We use a different notation here, which results from some rescalings and is meant to stress the symmetric behavior of the system under the two supersymmetry transformations.},
\beq
d_{ABC} \left[ W^{\alpha\,B}\, W_\alpha^C \ - \ 2\, \Phi^B \left( \frac{1}{4} \ \overline{D}^{\,2}\, \overline{\Phi}^C \ - \ m^C \right)\right] \ = 0 \ , \label{2_1_constraint_multiple}
\eeq
which rests on the totally symmetric coefficients $d_{ABC}$ and gives rise to multi--field Abelian extensions of the Born--Infeld model.
There is however a different non--linear limit, which is interesting in its own right, where both supersymmetries are broken in the presence of a vector multiplet, and in ${\cal N}=1$ superspace one can address it via suitably constrained superfields.

In this paper we would like to supplement the available list of superfield constraints with a new entry describing the ${\cal N}=2 \to {\cal N}=0$ breaking. It is a simple cubic modification of eq.~\eqref{2_1_constraint},
\beq
{\cal C}_3 \ : \ \Phi\, W^\alpha\, W_\alpha \ - \ \Phi^2 \left( \frac{1}{4} \ \overline{D}^{\,2}\, \overline{\Phi} \ - \ m \right) \ = \ 0 \ , \label{2_0_constraint}
\eeq
and the reader should note that a factor 2 is present in eq.~\eqref{2_1_constraint}, but not in eq.~\eqref{2_0_constraint}.

In order to justify the emergence of this new constraint, let us begin by recovering the $2 \to 1$ constraint \eqref{2_1_constraint} of \cite{BG} starting from the nilpotency condition
\beq
\Phi^2 \ = \ 0 \ . \label{phi2_constr}
\eeq
Its variation under the second supersymmetry of eqs.~\eqref{second_SUSY} yields a new constraint,
\beq
\Phi\, W_\alpha \ = \ 0 \ , \label{phiW_constr}
\eeq
and one more variation yields finally the constraint \eqref{2_1_constraint}, together with an additional term involving the derivative of the ``primary" constraint of eq.~\eqref{phi2_constr}. The reason behind this pattern is that the three constraints of eqs.~\eqref{2_1_constraint}, \eqref{phi2_constr} and \eqref{phiW_constr} build the chiral ${\cal N}=2$ constraint~\cite{RT}
\beq
{\cal X}^2 \ = \ 0 \ , \label{chi2}
\eeq
where
\beq
{\cal X}\ = \ \Phi \ + \ i\, \sqrt{2} \ {\tilde \theta}^{\alpha}\, W_{\alpha} \ - \ {\tilde \theta}^{\,2} \left( \frac{1}{4}\ {\bar D}^2 \, \bar \Phi \ - \ m\right) \ . \label{superspace_chi}
\eeq

In a similar fashion, starting from the cubic counterpart of eq.~\eqref{phi2_constr},
\beq
\Phi^3 \ = \ 0 \ , \label{phi3_constr}
\eeq
the same steps lead to
\beq
\Phi^2 \, W_\alpha \ = \ 0 \ , \label{phi2W_constr}
\eeq
and finally to eq.~\eqref{2_0_constraint}, together with another term involving the derivative of eq.~\eqref{phi3_constr}, which are the three ${\cal N}=1$ superfield components of the chiral ${\cal N}=2$ cubic constraint
\beq
{\cal X}^3 \ = \ 0 \ . \label{chi3}
\eeq

The fact that two different constraints are available for the ${\cal N}=2$ vector multiplet reflects the existence of two available options for ${\cal N}=2$ supersymmetry, the partial breaking to ${\cal N}=1$ and the total breaking. As a matter of fact, the new constraint of eq.~\eqref{2_0_constraint} admits necessarily a $2 \to 1$ branch of solutions. This is evident in ${\cal N}=2$ language, given the form of eqs.~\eqref{chi2} and \eqref{chi3}, but it is instructive to look into this fact in ${\cal N}=1$ language. The lesson is that, although the two sides of eqs.~\eqref{2_1_constraint} and \eqref{2_0_constraint} involve different relative coefficients, if eqs.~\eqref{phi2_constr} and \eqref{phiW_constr} hold, which is the case for the solution of \cite{BG}, the cubic constraint \eqref{2_0_constraint} also does, since both sides vanish. As we shall see, however, away from a singular corner the solution of ${\cal C}_3$ implements the \emph{total breaking of supersymmetry}. Let us also stress that the setting that are exploring for the total breaking  differs from the one captured by extensions of the ${\cal N}=2$ Volkov--Akulov Lagrangian (see \cite{dallfar} for a recent comprehensive discussion): when starting from an ${\cal N}=2$ vector multiplet, the resulting low--energy dynamics involves in fact a vector field $A_\mu$, in addition to the expected goldstini.

We have convinced ourselves that the cubic constraint of eq.~(\ref{2_0_constraint}) is the only possible generalization of the quadratic constraint (\ref{second_SUSY}), while all higher-order options that are apparently available do not yield new independent solutions. This phenomenon has a counterpart in the ${\cal N}=1$ Volkov--Akulov multiplet, where starting from a chiral superfield $X$ the only consistent solution~\cite{lledo} of the cubic constraint $X^3=0$ also solves the quadratic constraint $X^2=0$ of~\cite{nonlinear,ks}~\footnote{When several chiral multiplets are present, higher--order constraints can also arise in the IR from microscopic models, with solutions that differ from the standard one of the quadratic constraint $X^2=0$~\cite{dud2011}.}.

As we have anticipated, away from a singular corner the cubic constraint of eq.~\eqref{2_0_constraint} describes the total breaking of ${\cal N}=2$ supersymmetry. The scalar field $Z$ is then determined solving its $\theta^2$--component,
\begin{eqnarray}
&& \left[D^2 + 2\,F({\overline F}+ m)  - \frac{1}{2}\,F_{\mu \nu}\,F^{\mu \nu} -  \frac{i}{2} \,F_{\mu \nu}\, {\widetilde F}^{\mu \nu}
- 2\, i\, \psi \,\sigma^\mu \,\partial_\mu {\bar \psi} - 2 \,i
\, \lambda \,\sigma^\mu \,\partial_\mu {\bar \lambda}
+ Z \Box {\bar Z} \right] Z \ \nonumber \\
&& = \ \left({\overline F}\,+\,m\right) \psi\,\psi \ - \ i\,\sqrt{2} \,\psi
\left(D\ + \ i\, \sigma^{\mu \nu} F_{\mu \nu}\right) \lambda \ +\  F\,\lambda \, \lambda \ , \label{f5intro}
\end{eqnarray}
by a number of iterations, somehow
in analogy with the simpler procedure followed for the quadratic Volkov--Akulov constraint of \cite{nonlinear,ks}. One can verify that the lower components of eq.~\eqref{2_0_constraint} are identically satisfied by the solution of eq.~\eqref{f5intro}.

This is the generic state of affairs, but there is indeed a singular corner.
When the parameters in the Lagrangian realizing the ${\cal N}=2 \to {\cal N}=0$ breaking make the combination $D^2 + 2\,F({\overline F}+ m)$, and thus the scale $\Lambda$ of eq.~\eqref{scale}, vanish in the vacuum, this branch disappears but the partial breaking ${\cal N}=2 \to {\cal N}=1$ solution is still available, and makes both sides of eq.~\eqref{f5intro} vanish. In this singular corner, one can recover the expected ${\cal N}=2 \to {\cal N}=1$ supersymmetric Born--Infeld structure, since one is then solving effectively the quadratic constraint of \cite{BG}. Generically, however, the Lagrangian that we shall associate to eq.~\eqref{2_0_constraint} does not embody a Born-Infeld structure. This behavior is reminiscent of the ${\cal N}=1 \to {\cal N}=0$ breaking in the presence of a vector multiplet discussed in \cite{ks}. Still, we shall see that one can introduce vector self--couplings via additional terms involving integrals over the full ${\cal N}=2 $ superspace.

The contents of this paper are as follows. After identifying the new constraint in Section \ref{sec:identif_constraint} and after solving it in the generic case in Section \ref{sec:constraint}, in Section \ref{sec:superspace} we rephrase matters in ${\cal N}=2$ language and then turn to examine in detail, in Section \ref{sec:nonlinearl}, the construction starting from the simplest choice of Lagrangian. In Section \ref{sec:nonlinearlBI} we examine its singular ${\cal N}=2 \to {\cal N}=1$ corner, in order to elaborate on the Born--Infeld action in the presence of a Fayet--Iliopoulos term. In Section \ref{sec:BIl} we discuss how to enforce a nonlinear structure, in particular of the Born--Infeld type, in the generic ${\cal N}=2 \to {\cal N}=0$ case. In Section \ref{sec:UV} we provide a microscopic derivation of the cubic constraint \eqref{2_0_constraint}, along the lines of \cite{ks}, before concluding in Section \ref{sec:concl} with some considerations related to the ${\cal N}=2$ tensor multiplet.

%
\section{A simple way of identifying the constraint \eqref{2_0_constraint}
} \label{sec:identif_constraint}
In order to justify the form of the new constraint \eqref{2_0_constraint} in simple and intuitive terms, let us confine momentarily our attention to the case of constant fields, while taking into account that the ${\cal N}=2 \to {\cal N}=0$ breaking ought to identify the two fermions of the ${\cal N}=2$ vector multiplet, $\psi$ and $\lambda$, as goldstini for the two broken supersymmetries. In analogy with the more familiar ${\cal N}=1 \to {\cal N}=0$ case \cite{nonlinear,ks}, it is then natural to demand that the constraint eliminate the scalar $Z$ in terms of bilinear combinations of the two fermions. These would also depend, in general, on the gauge field strength, but if all fields are constant the gauge field strength vanishes and one is left with a generic combination of Fermi bilinears,
\beq
Z \ = \ a\, \psi\, \psi \ + \ b\, \psi\,\lambda \ +\ c\, \lambda \,\lambda \ , \label{f1}
\eeq
pretty much along the lines of \cite{nonlinear,ks}. The three constants $a,b,c$ are then to be chosen so that both sides behave identically under the two supersymmetries, which in this limit reduce to their non--derivative portions,
\bea
\delta \, Z &=&   \sqrt{2}\, \left( \epsilon_1\,\psi \ + \ \epsilon_2\,\lambda\right) \ ,  \nonumber \\
\delta \, \psi &=&  \sqrt{2}\, \epsilon_1\,F \ + \ i\, \epsilon_2\,D   \ , \nonumber \\
\delta \, \lambda &=&  i\,\epsilon_1\,D \ + \ \sqrt{2} \left(m \ + \ \overline{F}\right) \epsilon_2 \ .
\eea
One is thus led to an over-determined linear system for $a$, $b$ and $c$, which admits nonetheless the unique solution
\begin{equation}
 Z \ = \ \frac{\left({\overline F}\ +\ m\right) \psi \,\psi \ - \ i \ \sqrt{2} \,D \,\psi\, \lambda \, +\, F\, \lambda\, \lambda}{\left[  D^2\  +\  2\,F\left({\overline F} \,+ \,m\right)  \right]} \ . \label{f4}
\end{equation}
This expression is well defined if the denominator, whose absolute value already appeared in eq.~\eqref{MdagMintro2} and is related to the scale $\Lambda$ of eq.~\eqref{scale}, does not vanish. When this is case, both supersymmetries are broken.

For constant field configurations, eq.~\eqref{f4} solves indeed the $\theta^2$-- portion of the constraint \eqref{2_0_constraint}. On the other hand, its lower components reduce to
\bea
&& Z\ \lambda\,\lambda \ - \ Z^2 \left( {\overline F} \ + \ m \right) \ = 0 \ ,\nonumber \\
&& \lambda\,\lambda\ \psi_\alpha \ + \ i\,\sqrt{2}\,Z\,D \ \lambda_\alpha \ - \ 2\,Z \, \psi_\alpha \left( {\overline F} \ + \ m \right) \ = \ 0 \ , \label{lowertheta20}
\eea
and are solved identically by the composite scalar field $Z$ of eq.~\eqref{f4}, in close analogy with what happens for the ${\cal N}=1 \to {\cal N}=0$ constraint of \cite{nonlinear,ks}. This result should be contrasted with the behavior of the ${\cal N}=2 \to {\cal N}=1$ constraint of \cite{BG}, which works quite differently. Its components determine indeed, independently, the three fields of the chiral multiplet in terms of the vector multiplet.

For generic field configurations, expanding the superfields in eq.~\eqref{2_0_constraint} one can recover the terms in eq.~\eqref{f4}, together with others that contain derivatives and were neglected in the preceding argument. In particular, the complete $\theta^2$--component reads
\begin{eqnarray}
&& \left[D^2 + 2\,F({\overline F}+ m)  - \frac{1}{2}\,F_{\mu \nu}\,F^{\mu \nu} -  \frac{i}{2} \,F_{\mu \nu}\, {\widetilde F}^{\mu \nu}
- 2\, i\, \psi \,\sigma^\mu \,\partial_\mu {\bar \psi} - 2 \,i
\, \lambda \,\sigma^\mu \,\partial_\mu {\bar \lambda}
+ Z \Box {\bar Z} \right] Z \ \nonumber \\
&& = \ \left({\overline F}\,+\,m\right) \psi\,\psi \ - \ i\,\sqrt{2} \,\psi
\left(D\ + \ i\, \sigma^{\mu \nu} F_{\mu \nu}\right) \lambda \ +\  F\,\lambda \, \lambda \ . \label{f5}
\end{eqnarray}
This is a seemingly complicated non--linear differential equation for $Z$. However, the presence of anticommuting terms makes it possible to solve it, as we can now show.


\section{Explicit solution of the constraint \eqref{2_0_constraint}} \label{sec:constraint}

We can now describe how to obtain an explicit solution of the ${\cal N}=2 \to {\cal N}=0$ constraint of eq.~\eqref{2_0_constraint}. To this end, let us begin by noticing that eq.~\eqref{f5} is of the form
\beq
\left( a \ + \ Z \,\Box\, \overline{Z}\right) Z \ = \ b \ , \label{eqz}
\eeq
with
\bea
a &=& D^2 + 2\,F({\overline F}+ m)  - \frac{1}{2}\,F_{\mu \nu}\left(F^{\mu \nu} +  {i} \, {\widetilde F}^{\mu \nu}\right)
- 2\, i\, \psi \,\sigma^\mu \,\partial_\mu {\bar \psi} - 2 \,i
\, \lambda \,\sigma^\mu \,\partial_\mu {\bar \lambda}
\ ,  \label{sol:a}  \\
b &=& \left({\overline F}\,+\,m\right) \psi\,\psi \ - \ i\,\sqrt{2} \,\psi
\left(D\ + \ i\, \sigma^{\mu \nu} F_{\mu \nu}\right) \lambda \ +\  F\,\lambda \, \lambda \ . \label{sol:b}
 \eea
Notice that $b$ is a combination of fermionic bilinears. Its two important properties for us are
\bea
b^2 &=& a_0\ \left(\psi\,\psi\right)\left(\lambda\,\lambda\right)\ ,  \nonumber \\
b^3 &=& 0 \ , \label{constraintsb}
\eea
where $a_0$ denotes the purely bosonic portion of $a$:
\beq
a_0 \ = \ D^2 \ + \ 2\,F({\overline F}+ m) \ -\  \frac{1}{2}\,F_{\mu \nu}\,F^{\mu \nu} \ -\  \frac{i}{2} \,F_{\mu \nu}\, {\widetilde F}^{\mu \nu} \ .
\eeq

A first direct consequence of the preceding relation is a very simple solution for $Z^2$,
\beq
Z^{\,2} \ = \ \frac{\left(\psi\,\psi\right)\left(\lambda\,\lambda\right)}{a_0} \ \ .
\eeq
This result holds since all other contributions vanish identically due to the anticommuting nature of the two--component spinors $\psi$ and $\lambda$.

Solving for $Z$, however, is a more difficult task that requires a few steps. To begin with, eq.~\eqref{eqz} can be reduced to the \emph{linear} constraint
\beq
Z \ = \ \frac{b}{a} \left( 1 \ - \ \frac{b}{a^2} \ \Box \,\overline{Z} \right) \ , \label{eqZalmostlin}
\eeq
making use of eqs.~\eqref{constraintsb}, and this result reflects another simple relation:
\beq
\frac{b}{a} \ Z \ = \ \frac{b^2}{a^2} \ .
\eeq
One can do better, however, since combining eq.~\eqref{eqZalmostlin} with its complex conjugate leads to a linear differential equation for $Z$ alone,
\beq
\left[1 \ - \ \frac{b^2}{a^3} \ \Box \left( \frac{\overline{b}^2}{\overline{a}^3} \ \Box \right) \right] Z \ = \ \frac{b}{a} \left[ 1 \ - \ \frac{b}{a^2} \ \Box \left( \frac{\overline{b}}{\overline{a}} \right)\right] \ ,
\eeq
which can be solved by successive iterations. To this end, it is convenient to define
\beq
u \ = \ \frac{b^2}{a^3} \ = \ \frac{\left(\psi\,\psi\right)\left(\lambda\,\lambda\right)}{a_0^2} \ , \qquad {\cal O} \ = \ \frac{b}{a}  \ - \ u \ \Box \left( \frac{\overline{b}}{\overline{a}} \right) \ ,
\eeq
and to notice that eq.~\eqref{constraintsb} has some clear implications, including
\beq
u \ \partial_\mu \, b \ = \ 0 \ , \qquad \partial_\mu \, u \ \partial_\nu \, u \ = \ 0 \ , \qquad u \ \partial_\mu\, \partial_\nu \, u \ = \ 0 \ ,
\eeq
since a product of two $u$'s vanishes unless at least four derivatives are distributed among them.

Taking these restrictions into account, one can show that the series terminates after one step, so that the complete solution reads
\beq
Z \ = \ {\cal O} \ + \ u\,\Box \left( \overline{u}\, \Box {\cal O}\right)  \ . \label{sol1}
\eeq
Let us also note, for future reference, that up to total derivatives that we shall ignore in the action,
\beq
{\overline Z}\, \Box \, Z \ = \ {\overline {\cal O}} \, \Box \, {\cal O} \ + \  2 \left( {\overline u} \, \Box \, {\cal O}\right) \, \Box \, \left({u} \, \Box \, {\overline {\cal O}}\right)  \ . \label{ZboZnl}
\eeq

The two discrete symmetries
\bea
&&\psi \ \to \ i\,\lambda \ , \quad \lambda \ \to \ \pm\,i\, \psi \ , \quad F\ \to \ - \ \bigl({\overline F} \ + \ m\bigr)  \ , \quad Z \ \to \ Z\nonumber \\ &&{\overline F} \ \to \ - \ \bigl({F} \ + \ m\bigr) \ ,
\quad D \ \to \ \mp \ D \ , \quad A_\mu \ \to \ \pm \ A_\mu   \label{sol2}
\eea
corresponding to the upper or lower signs above, are similar to the discrete R--symmetry described by Fayet in \cite{fayet} and can flip, independently, the signs of $D$ or $A_\mu$ in the solution for $Z$ of eq.~\eqref{sol1}. This property is very interesting: it implies that the gauge multiplet can be consistently removed from the Lagrangian that in Section \ref{sec:nonlinearl} we shall associate to the ${\cal N}=2 \to {\cal N}=0$ breaking and that, in the absence of the Fayet--Iliopoulos term, $D=0$ solves its equation of motion. Our construction therefore reduces, in this case, to the Volkov--Akulov model of \cite{nonlinear,ks}.


\section{${\cal N}=2$ superspace and the cubic constraint \eqref{2_0_constraint}} \label{sec:superspace}

In the superspace approach of \cite{grimmwess}, the ${\cal N}=2$ vector multiplet can be accommodated in doubly chiral superfield as
\beq
{\cal X}\ = \ \Phi \ + \ i\, \sqrt{2} \ {\tilde \theta}^{\alpha}\, W_{\alpha} \ - \ {\tilde \theta}^{\,2} \left( \frac{1}{4}\ {\bar D}^2 \, \bar \Phi \ - \ m\right) \ , \label{superspace1}
\eeq
where $\tilde \theta$ is a second fermionic superspace coordinate and $m$ is the (magnetic) deformation parameter responsible for the partial breaking of supersymmetry  \cite{APT,BG,RT}.  One can simply verify that in this formalism the constraint (\ref{2_1_constraint}) of \cite{BG} that encodes the partial breaking of supersymmetry is simply
\beq
{\cal X}^2 \ = \ 0  \ . \label{superspace2}
\eeq
On the other hand, the constraint (\ref{2_0_constraint}) that in this paper we are associating to the total ${\cal N}=2 \to {\cal N}=0$ supersymmetry breaking in a vector multiplet reads
\beq
{\cal X}^3 \ = \ 0  \ . \label{superspace3}
\eeq

The Lagrangian for an ${\cal N}=2$ vector multiplet rests on an integral over the chiral half of the ${\cal N}=2$ superspace, and reads
\beq
{\cal L} \ = \   {\rm Im} \int d^2 \theta \,d^2  \tilde \theta \ {\cal F} ({\cal X}) \ + \ {\rm Im} \, \int d^2 \theta \ e_c\ \Phi \ , \label{n2gsw}
\eeq
where $e_c$ is a complex charge and ${\cal F}({\cal X})$ is the ${\cal N}=2$ prepotential. Notice that the second term is also allowed, since it varies into a total derivative under the transformations in eq.~\eqref{second_SUSY}\footnote{A Fayet-Iliopoulos term is also allowed by ${\cal N}=2$ supersymmetry.}.
Computing the chiral ${\tilde \theta}$ integral leads to the formulation in ${\cal N}=1$ superspace,
\beq
{\cal L} \ = \ \int d^4 \theta \ {\cal K} (\Phi, \bar \Phi)\ + \ \int d^2 \theta \left[\frac{1}{4}\, f(\Phi) \ W^{\alpha}\, W_{\alpha}\ + \ {\cal W} (\Phi)\ + \ {\rm h.c.} \right] \ , \label{superspace11}
\eeq
where
\bea
&& {\cal K} \ = \ \frac{i}{2}  \left(   \Phi \ \frac{\partial \bar {\cal F}}{\partial \bar \Phi} \ - \   \bar \Phi \ \frac{\partial  {\cal F}}{\partial  \Phi} \right) \ , \nonumber \\
&& f \ =\ -\ {i} \  \frac{\partial^2  \cal F}{\partial \, \Phi^2}  \ , \quad {\cal W} \ = \ -\ \frac{i\, e_c}{2} \,\Phi \ -\ \frac{i\, m }{2}\ \frac{\partial \, \cal F}{\partial \, \Phi}    \ . \label{superspace4}
\eea

In general, ${\cal N}=2$ Fayet--Iliopoulos terms are given in terms of electric and magnetic fluxes $e_X$ and $m_X$, $(X=1,2,3)$, which are triplets of the $SU(2)$ R-symmetry. In ${\cal N}=1$ language, the complexified electric and magnetic charges $e_c=e_1-i\,e_2$ and $m_1+i m_2$ enter the superpotential, while the third components are ${\cal N}=1$ Fayet--Iliopoulos terms. Using the $SU(2)$ R-symmetry, one can always set $e_3=0$ and $m_2=m_3=0$, so that one is left with a complexified electric charge $e_c$ and a real magnetic charge $m$, as in the preceding expressions. This standard choice was made in \cite{BG,RT,STORA} to analyze the supersymmetric Born--Infeld system, and has the virtue of aligning the unbroken supersymmetry with the ${\cal N}=1$ superspace. This would not the case if one introduced a Fayet--Iliopoulos term, as in the original work of~\cite{APT}. We shall return to this issue in Section \ref{sec:nonlinearlBI}.

The quantity
\beq
\mu_X \ = \ \epsilon_{XYZ}\ e^Y\ m^Z
\eeq
determines the splitting of the two eigenvalues of the matrix ${\cal M}$ of eq.~\eqref{matrixMitro} that encodes the vacuum fermionic shifts, to which we shall return in the next section, according to
\beq
V_{\pm} \ = \ V \ \pm \ \left| \mu \right| \ ,
\eeq
and with the preceding choice of charges $\left| \mu \right|= \left|m\, e_2\right|$.

The ${\cal N}=2 \to {\cal N}=1$ case of \cite{BG} rests on the choice~\footnote{A microscopic description of the Born--Infeld Lagrangian would also require, as in \cite{RT,STORA}, a cubic term in the prepotential, but once the constraint is assumed this choice is sufficient.}
\beq
{\cal F} \ = \ \frac{i}{2}\ {\cal X}^{\,2} \ ,
\eeq
where ${\cal X}$ is subject to the constraint of \eqref{superspace2}. Alternatively one can use, to begin with, ${\cal F}=0$, and in both cases one is finally led to
\beq
{\cal L} \ = \ {\rm Im} \, \int d^2 \theta \ e_c\ \Phi \ ,
\eeq
consistently with the vanishing of the ${\cal N}=2$ prepotential on the constraint surface. The complete ${\cal N}=1$ supersymmetric Born--Infeld Lagrangian is thus recovered from the $F$--term.

On the other hand, in the ${\cal N}=2 \to {\cal N}=0$ case one must start with proper kinetic terms, since as we have seen the cubic constraint of eq.~\eqref{2_0_constraint} expresses the scalar field $Z$ of the Wess--Zumino multiplet in terms of fermionic bilinears. The simplest option is
\beq
{\cal F} \ = \ \frac{i}{2}\ {\cal X}^{\,2}  \ , \label{quadratic_prepotential}
\eeq
with ${\cal X}$ subject to the constraint of eq.~\eqref{superspace3}, and the resulting Lagrangian is
\beq
{\cal L} \, = \, \int d^4 \theta \ \bar{\Phi}\, \Phi \, + \, {\rm Im} \, \int d^2 \theta \ e_c\ \Phi \, + \, \left[ \frac{1}{4} \ \int d^2 \theta \ W^\alpha\,W_\alpha \, + \, \frac{m}{2} \ \int d^2 \theta \ \phi \, + \, {\rm h.c.}\right] \, . \label{action_complex_e}
\eeq

In conclusion, one can describe the partial or total breaking of ${\cal N}=2$ supersymmetry combining the prepotential of eq.~\eqref{quadratic_prepotential} with one of the two constraints
\bea
&& {\cal N}=2 \ \to\  {\cal N}=1 \ : \qquad {\cal X}^2 \ = \ 0 \ ,  \nonumber \\
&&  {\cal N}=2\  \to \ {\cal N}=0 \ : \qquad {\cal X}^3 \ = \ 0  \ . \label{superspace5_20}
\eea
The first choice yields the supersymmetric Born--Infeld Lagrangian \cite{CF,BG}, while the second leads to non--linear contributions that are confined to fermionic couplings, and thus to an ${\cal N}=2$ counterpart of the Volkov--Akulov
construction. The resulting model captures a total breaking of supersymmetry that leaves behind, at low energies, not only a pair of goldstini but also a gauge field. Finally, let us stress again that no higher (quartic, etc) superfield constraints are expected to yield independent consistent solutions.

\section{On the non--linear ${\cal N}=2 \to {\cal N}=0$ Lagrangian} \label{sec:nonlinearl}

The nature of the cubic constraints of eq.~\eqref{2_0_constraint}, which relate the scalar $Z$ to fermion bilinears, leads naturally to select the quadratic prepotential in eq.~\eqref{superspace5_20} for a model Lagrangian, in order to grant proper kinetic terms to all fields. We shall explore this type of system, which bears some similarities to the Volkov--Akulov model in \cite{nonlinear,ks}, in two contexts, first introducing a Fayet--Iliopoulos term $\xi$ and a real electric charge $e$ and then introducing only a complex electric charge $e_c$, as in the preceding section.

The first type of low--energy Lagrangian of interest is therefore
\beq
{\cal L} \ = \ \int d^4 \theta \, \bigl( {\overline \Phi}\, \Phi\ + \ \xi V \bigr) \ + \ \int d^2 \theta\, \left[\frac{1}{4}\  W^{\alpha}\, W_{\alpha}\ + \ \frac{1}{2} \,\left(m\,- \,{i e } \right) \Phi \right] \ + \ {\rm h.c.}  \ . \label{nll3}
\eeq
On the constraint surface of eq.~\eqref{f5}, which can generally be solved for $Z$, this Lagrangian
describes the ${\cal N}=2 \to {\cal N}=0$ breaking in the non--linear limit. Its component form is simply
\bea
 {\cal L}  &=&  - \ i \,\psi \sigma^{\mu} \partial_{\mu} {\bar \psi}\ - \  i \,\lambda \,\sigma^{\mu} \partial_{\mu} {\bar \lambda} \ - \ \frac{1}{4}\ F_{\mu \nu}\,F^{\mu \nu} \ +\ |F|^2 \nonumber \\&+& \frac{1}{2}\ (m\,-\,ie)\, F \ +\
\frac{1}{2}\ (m\,+\,ie)\, {\overline F} \ + \ \frac{1}{2}\ D^2\ + \ \frac{\xi}{2} \ D \ +\  {\bar Z}\, \Box\, Z \ .  \label{nll4}
 \eea
In this expression one ought to introduce the solution of the constraint (\ref{sol1}), which was given in eq.~\eqref{ZboZnl}, to then solve for the $F$ and $D$ auxiliary fields accordingly, again by successive iterations. These solutions would result in expansions of the type
\beq
F \ =\  - \ \frac{1}{2}\ (m\,+\,ie) \,+ \,  {\cal O} {\rm (4-Fermi)} \ , \qquad D \ =\  - \  \frac{\xi}{2}   \ +  \  {\cal O} {\rm (4-Fermi)}  \ .  \label{n115}
\eeq
For brevity, here we shall confine our attention to the lowest--order terms, which suffice to exhibit four--fermion interactions and contain already some interesting lessons, leaving aside the corrections that we left implicit in eqs.~\eqref{n115}. All in all, up to and including four--Fermi terms one is thus led to
\beq
 {\cal L}  \ = \  -\ \frac{2(e^2\,+\, m^2)\, + \,\xi^2}{8}  \ - \  i \,\psi\, \sigma^{\mu}\, \partial_{\mu} {\bar \psi} \ - \  i \,\lambda\, \sigma^{\mu} \,\partial_{\mu} {\bar \lambda} \ - \ \frac{1}{4} \,F_{\mu \nu}\,F^{\mu \nu}  \ +\
{\bar Z}_0 \,\Box\, Z_0 \ ,  \label{nl16}
\eeq
with
\beq
Z_0 =  \frac{(m\,+\,i\,e) ({\psi}\, {\psi} \,-\, { \lambda}\, { \lambda}) \ + \ {\sqrt{2}}\, { \psi} (i\,\xi \,+ \,2\, {\sigma}^{\mu \nu} F_{\mu \nu}) { \lambda} }{\frac{\xi^2}{2} \ - \ (m\,+\,i\,e)^2 \ - \ F_{\mu \nu}\,F^{\mu \nu} \  - \ {i}\, F_{\mu \nu} {\widetilde F}^{\mu \nu}  } \  . \label{nl17}
 \eeq

Some comments are now in order. First of all, the structure of eq.~\eqref{f5} implies, in view of the discrete symmetry of section \ref{sec:constraint}, that one can consistently  set to zero the entire the gauge multiplet $(A_{\mu}, \lambda, D)$. As a result, in this case the complete theory, including all higher--order corrections, would reduce to the Volkov--Akulov Lagrangian, up to field redefinitions as in \cite{ks}. Moreover $D=0$ solves the corresponding equation in the absence of a Fayet--Iliopoulos term. The next comment concerns the detailed nature of the non--linear couplings. Indeed, while the corrected Lagrangian of eqs.~\eqref{nl16} and \eqref{nl17} contains a number of higher--order gauge field couplings, these are always dressed by fermionic bilinears, along the lines of what was found in \cite{ks} for the ${\cal N}=1 \to{\cal N}=0$ case. There a Volkov--Akulov model was coupled to a gauge multiplet, while the constraint $X \,W_{\alpha} = 0$ eliminated the gaugino from the spectrum. The corresponding ${\cal N}=1 \to {\cal N}=0$  action was also nonlinear in the gauge field \cite{ks}, due to the presence of terms dressed by fermion bilinears. Therefore, both settings contain no analog of the Born--Infeld vector self--couplings that accompany in the ${\cal N}=2 \to {\cal N}=1$ case.

We can now address the relation of the electric and magnetic parameters of the model Lagrangian \eqref{nll3} to the supersymmetry breaking scale(s) and the vacuum energy. These data ought to reflect themselves in the vacuum energy entering eq.~\eqref{nl16}, and it suffices to confine the attention to the lowest--order terms in the supersymmetry variations of the two goldstini. These can be encoded, in general, in the two-by-two matrix
\beq
{\cal M} \ =  \ \left( \begin{array}{cc}
    \sqrt{2}\,F     &  i\, D  \\
  i \,D      &  \sqrt{2}\,\left({\overline F}+m\right)
\end{array} \right)\ , \label{matrixM}
\eeq
which enters their transformations
\beq
\begin{pmatrix}
    \delta \psi     \\
    \delta \lambda
\end{pmatrix}  =
\begin{pmatrix}
    \sqrt{2}\, F      & i D  \\
    i\, D       &  \sqrt{2} ({\overline F} \ +  \ m)
\end{pmatrix}
\begin{pmatrix}
    \epsilon_1   \\
    \epsilon_2
\end{pmatrix}  \ + \  \cdots \ ,
\eeq
and determines the two important quantities
\bea
{\rm tr} \left( {\cal M}^{\dagger}\, {\cal M} \right) &=& 2\bigl[\left|F\right|^2 \, +\, \left|F+m\right|^2 \,+\,D^2\bigr] \ , \nonumber \\
\Lambda^8 \ = \ \det \left( {\cal M}^{\dagger} \, {\cal M} \right) &=& \bigl|2\,F \left({\overline F} \ + \ m\right) \ + \ D^2 \bigr|^2  \ . \label{MdagM}
\eea

The first line in eq.~\eqref{MdagM} \emph{defines}, up to a normalization, the vacuum energy associated to the ${\cal N}=2$ algebra. Notice that this differs in general from the vacuum energy determined via $F$ and $D$ terms in ${\cal N}=1$ superspace, as stressed in \cite{APT,adft}. Moreover, the determinant in the second line of eq.~\eqref{MdagM} is related to the scale $\Lambda$ of eq.~\eqref{scale}. It defines the range of validity of the iterative solution described in Section~\ref{sec:constraint} and used here, and thus of the resulting low--energy effective field theory.

Taking into account the vacuum values in eq.~\eqref{n115}, in the vacuum of our Lagrangian \eqref{nl16} ${\cal M}$ becomes
\beq
{\cal M} \ =  \ \frac{1}{\sqrt{2}} \ \left( \begin{array}{ll}
    - \ \left(m\ +\ i\,e\right)     &  \ - \ i\, \frac{\xi}{\sqrt{2}}  \\
  - \ i \,\frac{\xi}{\sqrt{2}}       &  m\ +\ i\,e
\end{array} \right)\ , \label{matrixM2}
\eeq
and
\beq
{\cal M}^{\dagger}\, {\cal M} \ = \left(\frac{e^2+m^2}{2} \ + \ \frac{\xi^2}{4}\right)  \ I_{2} \ - \  \frac{\xi\,m}{\sqrt{2}}\ \sigma_2 \ . \label{nll8}
\eeq
Therefore
\beq
{\rm tr} \left( {\cal M}^{\dagger}\, {\cal M} \right) \ = \ 2\left[\left|F\right|^2 \, +\, \left|F+m\right|^2 \,+\,D^2\right] \ = \ e^2 \ +\ m^2 \ + \ \frac{\xi^2}{2} \ , \label{vacuum}
\eeq
while the eigenvalues are
\beq
V_\pm \ = \ \frac{1}{2} \left[e^2 \ + \ \left( m\, \pm \, \frac{1}{\sqrt{2}}\ \xi\right)^2\right]
\eeq
which coincide in the absence of a Fayet--Iliopoulos term, and consequently
\beq
\Lambda^8 \ = \ \det \left( {\cal M}^{\dagger} \, {\cal M} \right) \ = \ \frac{1}{4} \left[e^2 \ + \ \left( m\,+\, \frac{1}{\sqrt{2}}\ \xi\right)^2 \right]\left[e^2 \ + \ \left( m\,-\, \frac{1}{\sqrt{2}}\ \xi\right)^2 \right]  \ . \label{scales}
\eeq

In this example, eq.~\eqref{vacuum} is actually proportional to the vacuum energy obtained from $F$ and $D$ terms in eq.~\eqref{nl16}, since the matrix $M$ in eq.~\eqref{matrixM} is traceless, but we shall exhibit a counterexample shortly. Before proceeding, however, let us also notice that there are choices of parameters for which the determinant vanishes, and thus the $2 \to 0$ solution becomes singular, namely $e=0$, $m= \pm \frac{\xi}{\sqrt{2}}$. On the other hand, in the absence of a Fayet--Iliopoulos term the two eigenvalues of the matrix ${\cal M}^{\dagger} \, {\cal M}$ coincide and do not vanish for nonzero choices of $e$ and $m$. At the singular point, the scale $\Lambda$ vanishes and the whole construction is to be reconsidered, since the expansion in Section \ref{sec:constraint} become singular. The Born--Infeld branch of the solutions of eq.~\eqref{f5}, which makes both sides of it vanish, continues to be available at these special points of parameter space. We shall return to it shortly.

There is an alternative way of describing this type of system, via the Lagrangian of eq.~\eqref{action_complex_e} where the Fayet--Iliopoulos term ($D$-term) was traded, via an $SU(2)$ R-symmetry rotation, for a complexified charge $e_c=e_1-i e_2$ in the superpotential ($F$-term), as in Section \ref{sec:superspace}. Retracing the preceding steps one would then be led to
\beq
{\cal M} \ =  \ \frac{1}{\sqrt{2}} \ \left( \begin{array}{cc}
    - \ \left(m\ +\ e_2\ + \ i\,e_1\right)     &  0  \\
  0       &  m\ - \ e_2 +\ i\,e_1
\end{array} \right)\ ,
\eeq
and therefore to
\beq
{\cal M}^{\dagger}\, {\cal M} \ = \frac{e_1^2\,+\,e_2^2\,+\,m^2}{2}  \ I_{2} \ +\  {m\,e_2} \ \sigma_3 \ . \label{nll82}
\eeq
Now ${\cal M}$ is not traceless, and so one is readily confronted with the subtlety that we had anticipated: the ${\cal N}=2$ vacuum energy, proportional to
\beq
{\rm tr} \left({\cal M}^{\dagger}\, {\cal M} \right) \ = \ e_1^2\ +\ e_2^2\ +\ m^2 \ , \label{MMdagcomplex}
\eeq
differs from the value that would be determined by the ${\cal N}=1$ $F$ and $D$ terms. On the other hand, eq.~\eqref{MMdagcomplex} is the correct value expected from the ${\cal N}=2$ Ward identity, while the ${\cal N}=1$ superspace formula does not take into account all contributions. The difference is conceptually important, although it only concerns a constant shift.

In general there are again two distinct eigenvalues,
\beq
V_\pm \ = \ \frac{1}{2} \left[e_1^2 \, +\, (m\, \mp \, e_2)^2 \right] \ ,
\eeq
so that
\beq
\det \left( {\cal M}^{\dagger}\, {\cal M} \right) \ = \ \frac{1}{4} \ \left[e_1^2 \, +\, (m+e_2)^2 \right] \  \left[ e_1^2 \, +\, (m-e_2)^2 \right]\ .
\eeq
The two eigenvalues coincide if $m\,e_2=0$, and therefore for the generic ${\cal N}=2 \to {\cal N}=0$ branch one can confine the attention to the case $e_2=0$.

In conclusion, the Lagrangian
\beq
{\cal L} \ = \ \int d^4 \theta \ {\overline \Phi}\, \Phi \ + \ \int d^2 \theta\, \left[\frac{1}{4}\  W^{\alpha}\, W_{\alpha}\ + \ \frac{1}{2} \,\left(m\,- \,{i e_c } \right) \Phi \right] \ + \ {\rm h.c.}  \ , \label{nll32}
\eeq
to which eq.~\eqref{nll3} reduces in the absence of a Fayet--Iliopoulos term, when combined with eq.~\eqref{f5} embodies a non--linear realization where ${\cal N}=2$ supersymmetry is fully broken. The resulting couplings are fully determined by the solution for $Z$ in Section~\ref{sec:constraint} and by the consequent iterative solution for the auxiliary field $F$, while $D$ vanishes.
If $e_1=0$ and $m=\mp e_2$ the system with a complex electric coupling also degenerates, the iterative solution of eq.~\eqref{f5} becomes singular since $\Lambda$ vanishes, and one is led again to the Born--Infeld branch, for which  its sides both vanish.

\section{Born--Infeld revisited} \label{sec:nonlinearlBI}

It is instructive to retrace the steps in \cite{RT}, enforcing the quadratic constraint of eq.~\eqref{2_1_constraint} while also adding a Fayet--Iliopoulos term. The theory would thus involve, to begin with, four parameters, the complex electric charge $e_c$, the scale $m$ entering the supersymmetry transformations and the constraint~\eqref{2_1_constraint} and the Fayet--Iliopoulos coefficient $\xi$. On the quadratic constraint, however, the Lagrangian reduces to
\beq
 {\cal L}  \ = \  -\  \frac{i}{2} \ e_c \, F \ +  \frac{i}{2} \ e_c^\star \, {\overline F} \ + \ \frac{\xi}{2} \ D \ ,  \label{nl16bi}
\eeq
with $e_c=e_1-ie_2$ as above, and it is instructive to work out its bosonic terms in detail. To this end, one first solves for the auxiliary field $F$ from
\beq
D^2 + 2\,F({\overline F}+ m)  - \frac{1}{2}\,F_{\mu \nu}\,F^{\mu \nu} -  \frac{i}{2} \,F_{\mu \nu}\, {\widetilde F}^{\mu \nu} \ = \ 0 \ , \label{biconstrsol}
\eeq
which gives
\beq
F \ = \ -\ \frac{m}{2} \left[1 \ - \ \sqrt{1 \ -\  \frac{2D^2}{m^2} \ + \ \frac{1}{m^2}\ F_{\mu\nu}\,F^{\mu\nu} \ - \ \frac{1}{4 m^4} \left( F\cdot {\widetilde F}\right)^2} \right] \ + \ \frac{i}{4m} \ F_{\mu\nu}\, {\widetilde F}^{\mu\nu} \ ,
\eeq
The resulting Lagrangian in eq.~\eqref{nl16bi}, which now reads
\bea
 {\cal L}  &=& \frac{e_1}{4\,m} \ F\cdot {\widetilde F} \ + \ \frac{\xi}{2}\ D \nonumber \\ &+& \frac{m \, e_2}{2} \left[ 1 \ - \ \sqrt{1 \ -\  \frac{2\,D^2}{m^2} \ + \ \frac{1}{m^2}\ F_{\mu\nu}\,F^{\mu\nu} \ - \ \frac{1}{4 \, m^4} \left( F\cdot {\widetilde F}\right)^2} \right] \ ,
\eea
is then to be extremized with respect to $D$, which gives
\beq
D \ = \ - \ \frac{\xi\,m}{2\,e_2\,\sqrt{1 \ + \ \frac{\xi^2}{2\,e_2^2}}} \ \sqrt{1 \ + \ \frac{1}{m^2}\ F_{\mu\nu}\,F^{\mu\nu} \ - \ \frac{1}{4\, m^4} \left( F\cdot {\widetilde F}\right)^2} \ .
\eeq

As a result, the Lagrangian in eq.~\eqref{nl16bi} can be finally cast in the form
\bea
 {\cal L}  &=& \frac{e_1}{4\,m} \ F\cdot {\widetilde F} \ - \ \frac{m\,e_2}{2} \left[\sqrt{1 \ + \ \frac{\xi^2}{2\,e_2^2}} \ - \ 1 \right]\nonumber \\ &+& \frac{m \, e_2}{2} \ \sqrt{1 \ + \ \frac{\xi^2}{2\,e_2^2}} \ \left[ 1 \ - \ \sqrt{1 \ + \ \frac{1}{m^2}\ F_{\mu\nu}\,F^{\mu\nu} \ - \ \frac{1}{4\, m^4} \left( F\cdot {\widetilde F}\right)^2} \right] \ .
\eea
The Maxwell kinetic term contained in this expression is not in conventional form, but this can be recovered letting
\beq
{F}_{\mu\nu} \ \longrightarrow \ F_{\mu\nu} \left[\frac{m}{e_2\ \sqrt{1 \ + \ \frac{\xi^2}{2\,e_2^2}}} \right]^\frac{1}{2} \ , \label{rescaling}
\eeq
which turns the Lagrangian into
\bea
 {\cal L}  &=& \frac{\theta}{4} \ F\cdot {\widetilde F} \ - \ \frac{m\,e_2}{2} \left[\sqrt{1 \ + \ \frac{\xi^2}{2\,e_2^2}} \ - \ 1 \right]\nonumber \\ &+& \frac{\rho^{\,2}}{2} \ \left[ 1 \ - \ \sqrt{1 \ + \ \frac{1}{\rho^2}\ F_{\mu\nu}\,F^{\mu\nu} \ - \ \frac{1}{4\, \rho^4} \left( F\cdot {\widetilde F}\right)^2} \right] \ .
\eea
Here
\bea
\rho^{\,2} &=& {m \, e_2} \ \sqrt{1 \ + \ \frac{\xi^2}{2\,e_2^2}} \ , \nonumber \\
\theta &=& \frac{e_1}{e_2\ \sqrt{1 \ + \ \frac{\xi^2}{2\,e_2^2}}} \ ,
\eea
and the four original parameters $(m,e_1,e_2,\xi)$ have thus combined into three independent quantities: $\rho$, $\theta$ and the vacuum energy
\beq
{\cal E}_0 \ = \ \frac{m\,e_2}{2} \left[\sqrt{1 \ + \ \frac{\xi^2}{2\,e_2^2}} \ - \ 1 \right] \ . \label{E0misaligned}
\eeq
We would like to conclude this section with some comments on this result.

To begin with, the system describes an ${\cal N}=2 \to {\cal N}=1$ breaking, even in the presence of the Fayet--Iliopoulos term, since for one matter eq.~\eqref{biconstrsol} implies that, in the vacuum, the determinant of the matrix ${\cal M}$ of eq.~\eqref{matrixM} vanishes. Notice that in this case
\beq
{\cal M} \ =  \ \left( \begin{array}{cc}
 \frac{m}{\sqrt{2}} \ \left[ \ - 1 \ + \ \frac{1}{\sqrt{1 \ + \ \frac{\xi^2}{2\, e_2^2}}} \right]     &  - \ \frac{i\,\xi\, m}{2\,e_2\,\sqrt{1 \ + \ \frac{\xi^2}{2\, e_2^2}}}  \\
 - \ \frac{i\,\xi\, m}{2\,e_2\,\sqrt{1 \ + \ \frac{\xi^2}{2\, e_2^2}}}       &  \frac{m}{\sqrt{2}} \ \left[ 1 \ + \ \frac{1}{\sqrt{1 \ + \ \frac{\xi^2}{2\, e_2^2}}} \right]
\end{array} \right)\ , \label{matrixMBIxi}
\eeq
which also depends on $\xi$ and $e_2$, as the ${\cal N}=1$ energy ${\cal E}_0$ of eq.~\eqref{E0misaligned}. However, the ${\cal N}=2$ vacuum energy,
\beq
{\rm tr} \left({\cal M}^{\dagger}\, {\cal M} \right) \ = \ 2 \, m^2 \ ,
\eeq
only depends on $m$. There is, however, an apparent discrepancy between the
vacuum energy ${\cal E}_0$ and the value that one would expect from the algebra,
\beq
2 \, \left|F\right|^2 \ + \ D^2 \ = \ m^2 \left[ 1 \ - \ \frac{1}{\sqrt{1 \ + \ \frac{\xi^2}{2\, e_2^2}}} \right] \ ,
\eeq
which differs from it by a multiplicative factor. This is actually a nice illustration of a subtlety spelled out in~\cite{cfg87}: in higher--derivative theories Fermi fields can undergo a wavefunction renormalization, which must be properly taken into account in comparing these quantities! The proper rescaling is precisely due to the factor in eq.~\eqref{rescaling}, since the Fermi kinetic terms go along with the vector one, as can be seen from eq.~\eqref{f5}. Once this fact is taken into account, the complete parameter dependence of the ${\cal N}=1$ vacuum energy in eq.~\eqref{E0misaligned} is recovered from the algebra.

Why are we getting the peculiar non--vanishing result of eq.~\eqref{E0misaligned} for the ${\cal N}=1$ vacuum energy, which actually vanishes for $\xi=0$, in a theory with one unbroken supersymmetry? Simply because the introduction of $\xi$ has misaligned the unbroken supersymmetry with respect to the manifest ${\cal N}=1$ superspace~\cite{APT,adft}. $F$ and $D$ have both a non--vanishing vacuum value, and the Fayet--Iliopoulos term introduces no other effects, consistently with its redundancy. Indeed, in the ${\cal N}=2$ microscopic formulation of~\cite{APT}, as we have stressed above, $\xi$ could be rotated away by an $SU(2)$ R-symmetry transformation, reducing the two $SU(2)$ triplets of electric and magnetic Fayet--Iliopoulos terms to three independent parameters $(e_1,e_2,0)$ and $(m,0,0)$. This exercise can thus convey a few instructive lessons, and confirms in particular the relevance of the ${\cal N}=2$ vacuum energy for the dynamics of this class of models.

\section{An ${\cal N}=2 \to {\cal N}=0$ model with a Born-Infeld structure} \label{sec:BIl}

The  action for the ${\cal N}=2$ vector multiplet considered so far rests on an integral over the chiral half of the ${\cal N}=2$ superspace, combined with the triplet of Fayet--Iliopoulos terms, and reads
\beq
{\cal L}_0 \ = \   {\rm Im} \int d^2 \theta \,d^2  \tilde \theta \  \frac{i}{2}\ {\cal X}^2  \ + \  {\rm Im} \int d^2 \theta \ e_c\, \Phi  \ ,  \label{bi1}
\eeq
with ${\cal X}$ subject to the cubic constraint of eq.~\eqref{superspace5_20}. While in the partial ${\cal N}=2 \to {\cal N}=1$ breaking case the Lagrangian induced by the quadratic constraint in eq.~\eqref{superspace5_20} led to the Born---Infeld structure, here this is not true anymore. In the ${\cal N}=2 \to {\cal N}=0$ case there are no other contributions resting on chiral superspace integrals, but there are other terms that involve integrals over the full ${\cal N}=2$ superspace. The one of interest for us is
\beq
{\cal L}_1 \ = \   \int d^4 \theta \,d^4  \tilde \theta \   B(C,{\bar C} ) \ {\cal X}^2\ {\cal {\bar X}}^2 \  \ ,  \label{bi2}
\eeq
where the function $B$ is given in ~\cite{ketov},
\beq
B = \frac{1}{1 \ -\  \frac{1}{2}\, (C + {\bar C}) \ + \ \sqrt{1\,-\, C \,-\, {\bar C} \,+\, \frac{1}{4}\, (C \,-\, {\bar C})^2 }}   \ ,  \label{bi22}
\eeq
and
\beq
C \ =\  \frac{1}{2} \ {\bar D}^4 \,{\cal X}^2 \ .
\eeq
By construction, $  {\cal L}_0 + {\cal L}_1$
has a Born--Infeld structure, and without the cubic constraint of eq.~\eqref{superspace5_20} it would have ${\cal N}=2$ supersymmetry and would in principle describe a partial breaking  ${\cal N}=4 \to {\cal N}=2$. There is some controversy in the literature on the original construction
in \cite{ketov}, since the action lacks the manifest shift symmetry in the scalar $Z$ that would be expected for an interpretation in terms of brane position for a
D3 brane propagating in six dimensions~\cite{controversy}. In our case, however, the scalar $Z$ is eliminated by the constraint. Hence, the D-brane interpretation of our results, if any, should involve a fractional brane (\emph{i.e.} a brane confined to some orbifold fixed point, which would therefore lack the corresponding translational modes).
\section{A microscopic model leading to the constraint \eqref{2_0_constraint}} \label{sec:UV}

One can also exhibit a microscopic origin for the cubic constraint of eq.~(\ref{2_0_constraint}). As we shall see shortly there is an analogy, in this respect, with the ${\cal N}=1 \to {\cal N}=0$ case discussed in \cite{ks}, where the chiral constraint $X^2=0$ was recovered starting from a microscopic action of the type
\beq
{\cal L}_{\rho} \ = \ \int d^4 \theta \left[ {\overline X}\, X \ - \ \rho \, (\overline X X)^2 \right] \  +\  \left(\int d^2 \theta \ f \, X\ +\ {\rm h.c.}\right)   \ .  \label{uv1}
\eeq
For $\rho > 0 $, the quartic term gives a large mass to the scalar sgoldstino, and effectively imposes the constraint $X^2=0$ at scales below its mass.

In our   ${\cal N}=2 \to {\cal N}=0$ case, it can be shown that any Lagrangian based on integrals over a chiral half of the superspace, for any prepotential that is
polynomial in the superfield ${\cal X}$, generates a tachyonic eigenvalue in the scalar mass matrix, as already noticed in \cite{APT}. However, one can give rise to heavy scalars, and thus to large positive eigenvalues of the scalar mass matrix, adding integrals over the full  ${\cal N}=2$ superspace.
It turns out that the lowest--order polynomial leading to non-tachyonic scalar masses, in ${\cal N}=2$ superspace formalism,  is
\beq
{\cal L}_{\rho} \ =\  - \ \frac{\rho}{9} \int \ d^4 \theta\ d^4 {\tilde \theta}  \ {\cal X}^3 \ {\overline{\cal X}}^3   \ .  \label{uv2}
\eeq
In the standard  ${\cal N}=1$ superspace formalism, this expression translates into
\beq
{\cal L}_{\rho}\ = \ -\  \rho \int \ d^4 \theta  \left| \, \Phi\ W^\alpha\, W_\alpha \ -\   \frac{1}{2} \ \Phi^2 \left( \frac{1}{4} \ \overline{D}^{\,2}\, \overline{\Phi} \ - \ m \right)  \right|^2  \
= \- \ \rho\, | a\, Z \ - \ b|^2 \ +\ \cdots \  , \label{uv3}
\eeq
up to terms containing more derivative operators, which would not contribute to scalar masses, and where $a,b$ are defined in eqs.~(\ref{sol:a}) and (\ref{sol:b}).

The operator ${\cal L}_{\rho}$ therefore generates diagonal scalar masses, with $m_Z^2 = \rho |a|^2$, once vacuum values of the auxiliary fields induce the ${\cal N}=2 \to {\cal N}=0$ breaking of supersymmetry. At energies below these scalar masses
$m_Z$, the effective low energy theory would involve $(\psi,\lambda,A_{\mu})$, together with the auxiliary fields $(F,D)$ encoded in the vector multiplet
${\cal X}$, in the presence of the cubic constraint of eq.~\eqref{superspace5_20}.  One can understand this result noticing that in the large $\rho$ limit eq.~\eqref{uv3} implies for $Z$ the equation of motion $Z = \frac{b}{a}$. This is precisely the zero-momentum limit of the full solution  (\ref{sol1}), whose momentum--dependent part
is then completely determined by nonlinear supersymmetry, as prescribed by the cubic constraint of  eq.~(\ref{2_0_constraint}).

\section{A linear multiplet counterpart} \label{sec:concl}

The nonlinear realization of ${\cal N}=2$ supersymmetry based on the ${\cal N}=2$ tensor multiplet provides a natural counterpart of our construction. In ${\cal N}=2$ language \cite{WESS,RT} this multiplet is described by a real $SU(2)$ triplet of superfields $L^I$, or equivalently by a symmetric bi-spinor $L_{ab} = \left(\sigma^I \, \sigma^2\right)_{ab}\, L^I$, with $L_{ab}$ subject to the irreducibility constraint
\beq
D_{\alpha\,(a}\, L_{bc)} \ = \ 0  \qquad (a,b,c=1,2) \ ,
\eeq
where the $D_{\alpha\,a}$ are ${\cal N}=2$ covariant derivatives.

In ${\cal N}=1$ language, the field content translates into a pair of
antichiral superfields ${\widetilde X}=L^1 \,- \,i\,L^2$ and ${\widetilde \psi}_\alpha$, where ${\widetilde \psi}_\alpha = D_\alpha L^3$, with $L^3$ a ${\cal N}=1$ real linear multiplet ($D^{\,2}\,L^3={\overline D}^{\,2}\,L^3=0$). ${\widetilde \psi}_\alpha$ then plays the same role as $W_\alpha$ does for the vector multiplet, albeit with a reversed chirality, since
$D_\beta \,{\widetilde \psi}_\alpha = 0$.  The physical degrees of freedom contained in $L$ are a scalar $\phi$, an antisymmetric tensor $B_{\mu \nu}$ and a fermion. It is well known \cite{bg2} that this collection of fields can be a Goldstone multiplet for the partial breaking  ${\cal N}=2\to {\cal N}=1$.
The quadratic constraint of eq.~(\ref{2_1_constraint}), with the appropriate replacements,
\beq
{\widetilde \psi}^\alpha\, {\widetilde \psi}_\alpha \ = \ 2\, {\widetilde X} \left( \frac{1}{4} \ {D}^{\,2}\, {\overline {\widetilde X}} \ - \ m \right) \ , \label{c1}
\eeq
implies the two constraints
\beq
{\widetilde X}^2 \ = \ 0 \ , \qquad {\widetilde X} \ {\widetilde \psi}_\alpha \ = \ 0 \ . \label{c2}
\eeq
As a result, it is natural to consider the counterpart of eq.~\eqref{2_0_constraint}, the cubic constraint
\beq
  {\widetilde X}  \ {\widetilde \psi}^\alpha\,  {\widetilde \psi}_\alpha \ = \ {\widetilde X} ^2 \left( \frac{1}{4}\ D^{\,2}\, {\overline {\widetilde X}} \ - \ m \right) \ , \label{c3}
\eeq
which by ${\cal N}=2$ supersymmetry would also imply
\beq
{\widetilde X}^3 \ =\ 0 \  , \qquad  {\widetilde X}^2  \, {\widetilde \psi}_\alpha \ = \ 0 \ .
\eeq
Notice that ${\widetilde \psi}_\alpha$ admits the antichiral expansion
\beq
{\widetilde \psi}_\alpha \ = \ \lambda_\alpha \ + \ {\bar \theta}^{\dot \alpha} \,H_{\alpha \dot \alpha} \ + \ i \, {\bar \theta}^2  \, \partial_{\alpha  \dot \alpha} {\bar \lambda}^{\dot \alpha}  \ , \label{c4}
\eeq
where
\beq
H_{\alpha \dot \alpha} \  \sigma_{\mu}^{\alpha\,{\dot \alpha}} \ =\ \partial_{\mu} \, \phi \ + \ i\, H_{\mu}   \ . \label{c5}
\eeq
Here $H_{\mu} $ is the dual of the antisymmetric tensor field strength, which satisfies the condition $\partial^{\mu} H_{\mu} =0$. This theory has a gauge symmetry
for $B_{\mu \nu}$ and a shift symmetry in $\phi$.

In analogy with the previous case of the ${\cal N}=2$ vector multiplet, the cubic constraint of eq.~\eqref{c3} eliminates the scalar in $X$ and gives rise, generically, to a nonlinear Lagrangian describing the
total breaking of supersymmetry  ${\cal N}=2\to {\cal N}=0$. The resulting physical degrees of freedom would be the two goldstini, the scalar $\phi$ and the antisymmetric
tensor $b_{\mu \nu}$. Again, the simplest Lagrangian constructed as an integral over half of  ${\cal N} =2$ superspace does not accommodate
a Born--Infeld--like structure. As before, however, this could be induced by an additional term involving an integral over the full ${\cal N} =2$ superspace.
Notice that the constraints of eq.~(\ref{c2}) have also solutions that break spontaneously ${\cal N} =1$  to  ${\cal N} =0$, leaving behind one goldstino, the scalar $\phi$ and the antisymmetric tensor~\cite{gianguido}.
\vskip 24pt
\noindent {\large \bf Acknowledgements}

\noindent We  are grateful to  Q.~Bonnefoy and M.~Porrati for discussions and/or collaboration on related issues. ED is grateful to Scuola Normale and CERN-TH, and  AS is grateful to CPhT--\'Ecole Polytechnique, APC--U. Paris VII and CERN-TH for the kind hospitality extended to them while this work was in progress.   SF was supported in part by the CERN TH-Department and by INFN (IS CSN4-GSS-PI). AS was supported in part by Scuola Normale Superiore and by INFN (IS CSN4-GSS-PI).
\newpage
\setcounter{equation}{0}

\end{document}